# Learning variant product relationship and variation attributes from e-commerce website structures

Pedro Herrero-Vidal
Amazon.com
Seattle, WA, United States
phvidal@amazon.com

You-Lin Chen
Amazon.com
Seattle, WA, United States
cyoulin@amazon.com

Cris Liu
Amazon.com
Seattle, WA, United States
licris@amazon.com

Prithviraj Sen
Amazon.com
Sunnyvale, CA, United States
prithsen@amazon.com

Lichao Wang
Amazon.com
Seattle, WA, United States
lichwang@amazon.com

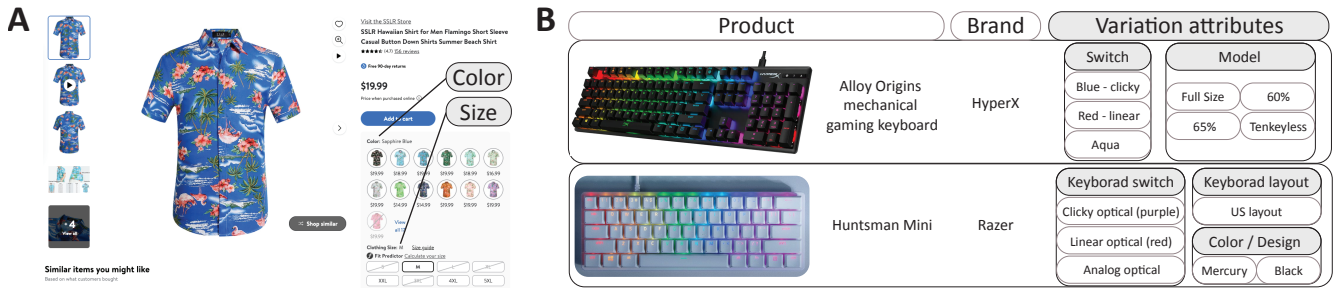

Figure 1: Variant product relationships in e-commerce sites. **A** E-commerce webpage screenshot showing a variant product listing example where the variation attributes, *color* and *size*, index the linked variant products on the same webpage. **B** Example showing how the same type of product, such as keyboards, can be associated to different *variation attributes* depending on the brand or listing website.

## ABSTRACT

We introduce VARM, <u>v</u>ariant <u>r</u>elationship <u>m</u>atcher strategy, to identify pairs of variant products in e-commerce catalogs. Traditional definitions of entity resolution are concerned with whether product mentions refer to the same underlying product. However, this fails to capture product relationships that are critical for e-commerce applications, such as having similar, but not identical, products listed on the same webpage or share reviews. Here, we formulate a new type of entity resolution in *variant product* relationships to capture these similar e-commerce product links. In contrast with the traditional definition, the new definition requires both identifying if two products are variant matches of each other *and* what are the attributes that vary between them. To satisfy these two requirements, we developed a strategy that leverages the strengths of both encoding and generative AI models. First, we construct a dataset that captures webpage product links, and therefore variant product relationships, to train an encoding LLM to predict variant matches for any given pair of products. Second, we use RAG prompted generative LLMs to extract variation and common attributes amongst groups of variant products. To validate our strategy, we evaluated model performance using real data from one of the world's leading e-commerce retailers. The results showed that our strategy outperforms alternative solutions and paves the way to exploiting these new type of product relationships.

## 1 INTRODUCTION

Entity resolution (ER) [4] is an important task in data integration whose goal is to determine whether two mentions refer to the same real-world entity. Industry practitioners and academic researchers have, for long, devised techniques to address ER in various domains, e.g. resolving social media handles [21], resolving products in e-commerce [8]. While ER usually refers to *exact* ER wherein two mentions are deemed to match each other if and only if each and every attribute of said mentions agree with each other, data integration in e-commerce entails addressing subtle but non-trivial versions of the basic ER task. Still, this traditional definition of entity resolution fails to capture product relationships that are critical for e-commerce catalogs.

As illustrated in the e-commerce site screenshot in Figure 1A, highly related products, but not identical products, are listed on the same webpage to facilitate the search. These consolidated webpage listings allow customers to look at the common product attributes while being able to easily choose amongst the different product variations, given by the *variation attributes*, such as color or size. Identifying these kind of product relationships not only improves e-commerce listings, but it can also be exploited for many other applications, such as review sharing or search deduplication. To capture this notion, we formulate a new type of entity resolution task for *variant product* relationships.

This new definition of variant products imposes additional considerations since it can vary across product types or even brands. For example, color and size are adequate variation attributes for clothing products, but for drinks the difference between products is their flavor or sugar content, or diagonal screen size for TVs (Fig. 4). Moreover, even for products of the same type, the specific variation attributes may change depending on the specific brand, such as keyboards from HyperX brand vary by *switch* and *model* whereas keyboards from Razer have *keyboard switch*, *keyboard layout* and *color/design* for variation attributes (Fig. 1B, Table 5).

In practice, identifying variant product relationships imposes two main challenges. Firstly, one has to establish if a given pair of products are variations of the same entity VARIANT MATCH or different products MISMATCH. Supervised methods based on encoding large language models (LLMs) are the current state of the art to establish if two products are identical EXACT MATCH, but it comes at the cost of collecting extensive labeled datasets for model training [15, 24], which are not readily available for variant matching, unlike exact matching. Secondly, one has to determine the variation attributes for the relevant set of variant products. While it could also be described as a supervised task, this would require learning thousands of variation attribute labels since they can vary by product type or brand, further increasing the need for labeled training data and the overall complexity of the task.

Here, we introduce a new strategy for variant relationship matching, VARM, that leverages the respective strengths of generative and encoding LLMs to overcome the challenges of identifying this new kind of product relationships. First, to capture variant product information, we construct a dataset that captures variant product pairs relationships given by products listed on the same webpage, which we used to train an encoding LLM to predict variant match relationships. Second, we use generative LLMs to predict variation attributes for groups of variant products, without requiring training data or being limited to a fixed set of variation attribute labels. To further provide the generative model with e-commerce information we used a retrieval-augmented generation (RAG) [5, 14] by providing context about products from similar product types and brands. Overall, this work presents the following main contributions:

- We introduce the novel variant product identification task.
- We formulate a novel method to identify variant products capitalizing on the information present in e-commerce webpages.
- We develop a strategy that uses the synergistic properties of encoding and generative LLMs to accurately predict variant matching products and variation attributes.
- We validate the model on three relevant datasets from e-commerce services.

## 2 RELATED WORK

Given that ER has been a topic of research for more than half a century, almost all major approaches of machine learning have been applied to solve it including supervised and unsupervised approaches [6]. More recently, the focus has shifted to deep learning including bespoke neural networks [17], pre-trained language models [16] and most recently, generative AI [18–20]. None of these works address variant matching, which is the focus of this work. Narayan et al. [18] show that GPT3 can perform entity resolution when provided with few-shot task demonstrations (in-context learning), Peeters and Bizer [19] show that the addition of entity matching rules to the prompt can help boost ChatGPT's matching quality and Peeters and Bizer [20] use GPT4 to provide explanations alongside match/mismatch predictions along with automatically dividing mispredictions into easy-to-understand error classes. More importantly, Peeters and Bizer show that despite recent advances, fine-tuning on sufficient labeled data can still outperform the best (zero-shot) generative AI entity resolution results. Following this result in our work, we fine-tune pre-trained large language models to generate variant match labels and to address the more challenging task of predicting variation attributes, we exploit world knowledge inherent in generative AI models.

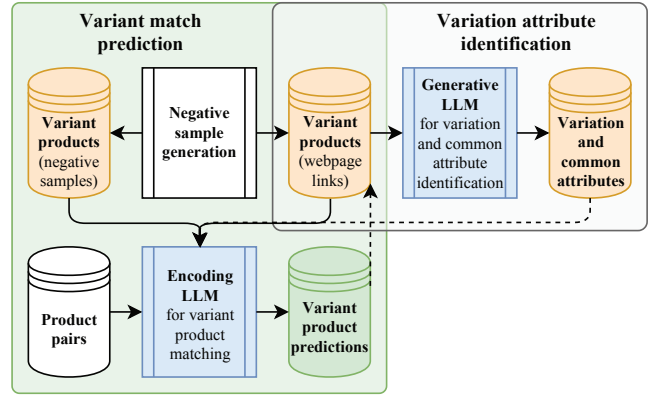

**Figure 2: VARM strategy schematic.** The variant product relationships present in webpages are exploited to construct a dataset with matching variant product pairs, which are then augmented to generate negative or mismatched examples for encoding LLM model training. The groups of variant products are also used to predict variation attributes using a generative AI that can also take RAG product information in the prompt. The trained models can be used to predict variant product relationships and attributes for any new pair of products.

## 3 METHODS

Our variant relationship matching, VARM, strategy can be largely divided into two main tasks (Fig. 2): 1) variant match prediction, and 2) identification of variation attributes.

### 3.1 Variant match prediction

Each product $p$ can be defined as a structured set of key-value pairs $p = \{(a_i, v_i)\}_{1 \leq i \leq k}$, where $a_i$ is the attribute name and $v_i$ is the attribute's value represented as text (see A.1). Given a product pair $(p_1, p_2)$ the encoding model aims to predict match label in a binary classification task.

DistilBERT was chosen as the encoding model given its competitive performance on product entity tasks [16, 20], but the strategy generalizes to alternative model choices. The model has 66M parameters, distilled from a 110M parameter teacher model, with 12 hidden layers, or transformer blocks and 768 attention heads [11, 22]. Given a pair of products, the product attributes from both products are first concatenated and tokenized into a single sequence of text tokens, to enable early fusion [27]. The two product descriptions are separated by a [SEP] token and padded and/or truncated to meet the 512 token input limitation, while ensuring that half the tokens came from each of the products in the pair. To provide the model with relevant product understanding, the model was first fine-tuned in e-commerce relevant tasks [10, 28]. In contrast to the off-the-shelf BERT, we will refer to version of the model pre-trained on the e-commerce dataset with an *ecom* tag.

To capture variant product relationships, a labeled dataset was generated capitalizing on the positive variant product links present in webpage listings, variation groups, and synthesizing negative samples by leveraging the positive links (see dataset details below). Model weights were fine-tuned to perform the variant product matching task in a supervised fashion by minimizing the cross-entropy loss function of a linear classification layer using this dataset. The training regime was limited to a single epoch with ADAM optimization and a $5e^{-6}$ learning rate without weight decay.

## 3.2 Variation attribute identification

For a set of products in a given variation group, we formulated the *variation attribute* estimation of VARM as a "Text-to-Text" task inspired by the recent success of generative AI [7, 13, 23]. Specifically, the input is structured as an instruction that encapsulates the text attributes for all the products belonging to a given variation group. Therefore, for each given variation group, we have a set of products $p_{1:k} \in \mathcal{P}$ with associated attributes $(a_{1:k,1:i}, v_{1:k,1:i})$, and instruction $\mathcal{I}$ that the generative model $f$ takes to predict attribute target class $c \in C$ as:

$$f : \{\mathcal{P}(a_{1:k,1:i}, v_{1:k,1:i}), \mathcal{I}\} \rightarrow C$$

where the set of class labels $C$ is limited to *common* and *variation* attributes. Note that we do not set a constraint to provide labels for all of the attributes nor to limit the output to structured attributes. As such, the model can identify variation attributes in the product attribute keys $a$ or values $v$. Still, we penalize the model when providing contradictory labels for the same attribute.

We formalize the model as both a zero-shot and few-shot learner utilizing an off-the-shelf LLM, Claude3 Haiku [1](see parameters settings in table 4). The zero-show formulation implicitly assumes that LLMs have been trained on massive amounts of language data and thus posses contextual understanding, given a correctly engineered prompt [12, 23]. For prompt engineering we combined Chain-of-Thought and instruction techniques, to predict all attributes labels for a given variation group [25, 29]. To mitigate the impact of this assumption, we provide product relevant information as part of the prompt using RAG [14]. For a given product pair of certain product type and brand, information about variation attributes is retrieved online from the webpage-linked products dataset. For the specific product type or brand, variation groups with products belonging to the same product type or brand were filtered, then the associated variation attributes were collected and structured into a list of unique variation attributes to be included in the prompt.

The detailed prompt provided to the model is structured as follows:

> ****** Variation attribute identification prompt ******
>
> You are an expert on products. The following list of products are the same entity but variations of each other. Some of the descriptors are unique to each product and some other are different across them. The attributes can be descriptors or keywords within the descriptor.
>
> <*if RAG*:><Usual different attributes for *product_type* products are {product_type_variation_attributes}. Usual different attributes for *brand* products are {brand_variation_attributes}.>
>
> Below are the products' descriptions: {variation_group_products}
>
> You need to complete the following tasks. Compare the details in the all products above and determine the attributes that are common and different across the products. If an attribute is "different", it cannot be "same".
>
> Respond: "Different:" followed by a list the attributes that are the different across them.
>
> Respond: "Same:" followed by a list the attributes that are same across them.
>
> Respond: "Reason:" explaining why or how attributes are different, for the different attributes.
>
> Do not add an explanation about the output format. Ensure the output is exclusively in JSON format. Return only a JSON block using double quotes in this format and in this order:
>
> {["Different": ["list different attributes"], "Same": [list same attributes], "Reason": [Reason why attributes are different]]} Do not return anything except a JSON. Always begin your output with: "{"

## 3.3 Datasets

The datasets used to develop and evaluate VARM models were[1]:

**Webpage-linked products** : dataset containing pairs of products listed on the same e-commerce webpage as illustrated in Fig. 1A. Since these pairs were presented together, we can assume them to belong to the same variation group and therefore are variant products, irrespective of the variation attribute(s). Using these positive relationships, we generated synthetic samples by shuffling product pairs. Exploiting our understanding of product relationships, we used an informed strategy to generate negative samples according to the following three buckets: hard negative samples coming from shuffling product pairs for another product from the same brand and product type, medium difficulty samples by shuffling product pairs for another product from the same product type but not brand, and easy samples by shuffling pairs for another product from different product type and brand. Text information from 2M product pairs and 168K variation groups is structured into product attributes, split into 70:30 training:evaluation sets with balanced class labels, while enforcing that products from the same variation groups are in the same data split.

---
[1]To the best of our knowledge, there are no publicly available datasets with variant product relationships.

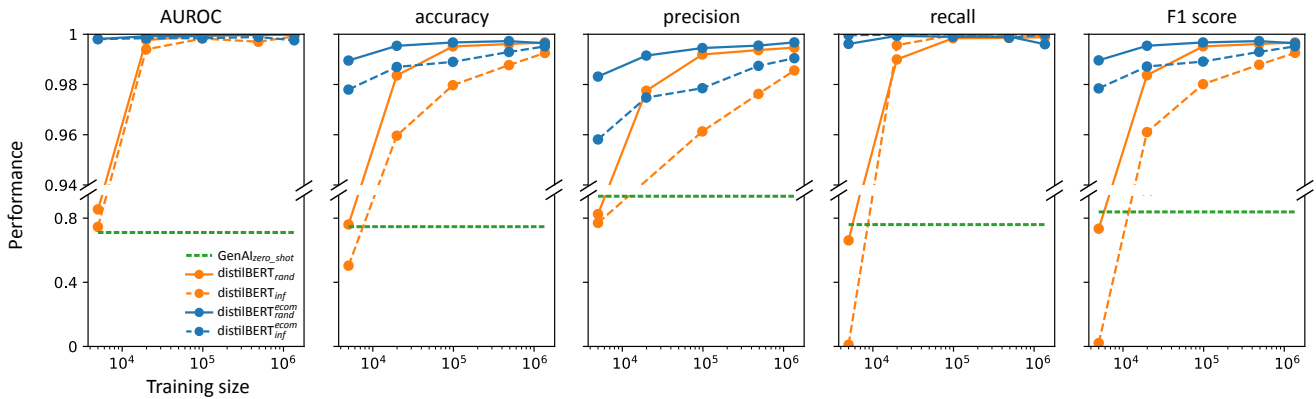

**Figure 3: Model performance as a function of training set size on the webpage linked products dataset.** Performance of genAI model $GenAI_{zero\_shot}$, off-the-shelf *DistilBERT* and DistilBERT pre-trained on e-commerce tasks $DistilBERT^{ecom}$, finetuned using random *rand* or informed *inf* negative sampling strategy.

**Expertly audited variant product pairs** : dataset containing 470 labeled pairs of products listed in different e-commerce websites and paired based on product similarity, representing a challenging dataset that mimics real use cases for variant product matching. The variant match labels were given by catalog experts taking webpage structure as ground truth.

**Expertly audited variation group attributes** : dataset with individual attributes from 10 variation groups, with 2-12 products each, labeled as *variation* or *common* depending on whether each attributes was different or the same across the products in the variation theme as determined by catalog human experts. The dataset contained 81 labeled attributes, with 35 and 46 labeled as variation and common, respectively.

## 4 RESULTS

### 4.1 VARM accurately learns variation matching product relationships from website structure

To validate our VARM strategy we evaluated the outputs of the different model components under different experimental conditions and compared its performance to state-of-the-art models performing the same tasks.

Firstly, since the webpage linked products dataset initially contains only positive variation match samples, we generated synthetic samples by shuffling product pairs. We used an informed strategy to generate the negative samples according to types of products and brands (see Datasets section in Methods). As control, we compared the performance of models trained in this dataset to that of a dataset containing random pair shuffles for negative sample generation. For each experiment we generated 1M negative samples to generate a label-balanced dataset with 2M pairs partition into 70:30 training:evaluation splits, while ensuring that products from the same variation group would be in a given split.

We fine-tuned encoding LLMs models on the aforementioned set of negative samples and the positive labels from the webpage-linked products dataset, starting from an off-the-shelf DistilBERT[2] model or a DistilBERT model pre-trained on e-commerce tasks, $distilBERT^{ecom}$. We also compared the performance of a zero-shot generative matching models using Claude on a held-out test split of the dataset (prompt details in A.3). While the generative model provides an above average performance without any training data, the fine-tuned models clearly outperformed it when trained on the task (Fig. 3).

To better understand the dependency of labeled data on model performance, we varied the amount of data used to fine-tune the encoding models. As expected, the performance increases monotonically with training set size. However, fine-tuning a LLM model from scratch, *DistilBERT*, requires tens of thousands of training examples to achieve similar performance to an equivalent model that was pre-trained on e-commerce tasks and further fine-tuned with a few thousand examples, $DistilBERT^{ecom}$. This is likely because the off-the-shelf model has to learn product representations alongside the variant matching task, whereas the $DistilBERT^{ecom}$ model only had to learn the new task boundaries. Moreover, the strategy used to generate the negative samples also impacts task learning rates. Fine-tuning with randomly generated samples $DistilBERT_{rand}$ results in performances comparable to those of training the same models with a hundred times less training data generated taking into consideration product type and brand dependencies $DistilBERT_{inf}$ (Fig. 3). Still, it is important to note that the difficulty of the task is influenced by the synthetically generated negative examples and may not be fully representative of the true distributions find in practice.

---
[2]https://huggingface.co/docs/transformers/en/model_doc/distilbert

| Model | AUROC | Accuracy | Precision | Recall | F1 score |
|---|---|---|---|---|---|
| $GenAI_{zero\_shot}$ | - | 74.68 | 92.01 | 75.98 | 83.89 |
| $distilBERT_{rand}$ | 68.17 | 82.12 | 91.78 | 88.23 | 89.55 |
| $distilBERT_{inf}$ | 69.98 | 86.81 | 92.41 | **99.99** | **92.93** |
| $distilBERT^{ecom}_{rand}$ | 89.71 | 85.74 | 97.90 | 87.74 | 91.44 |
| $distilBERT^{ecom}_{inf}$ | **90.61** | **87.44** | **98.39** | 90.68 | 92.61 |

**Table 1: Models performance comparison evaluated e-commerce relevant variant product dataset.**

To prove the validity of VARM's matching model component for practical e-commerce applications, we tested its performance in a dataset with variation product pairs across e-catalogs and expertly labeled. Given that the example pairs were sampled and not synthetically generated, they represent conditions that could be faced when tackling e-commerce tasks. Testing the generalization performance of the different models showed that all models can accurately estimate variant product relationships. Still, $DistilBERT_{inf}^{ecom}$ model pre-trained on e-commerce tasks before being fine-tuned on the variant matching tasks outperforms alternative models (Table 1).

## 4.2 VARM correctly classifies variation attributes

Identifying attributes that are common or vary across groups of variant products is critical for multiple applications. To assess VARM's ability to label attributes, we sampled 500 variation groups present in the webpage linked products dataset where the variation attribute is known. We tested the performance of the zero-shot generative AI model, $GenAI_{zero\_shot}$, prompted to solve this task and also when provided additional RAG information about variation attributes from other products of similar type and brand, $GenAI_{RAG}$. Both $GenAI_{zero\_shot}$ and $GenAI_{RAG}$ generate qualitative correct responses, with consistent explanations that take into consideration product type and brand (examples in A.2).

To get a quantitative estimate of the performance, we estimated the recall when predict structured variation attributes, to prevent penalizing for additionally found variation attributes that may not have a structured key. As baseline, we define a heuristic model that estimates variation attributes as the structured attributes that vary across more that 90% of the products in the variation group. All models can predict variation attributes above chance, with generative models outperforming heuristic-based methods. Moreover, providing additional context about products in the prompt, $GenAI_{RAG}$ further boosts attribute identification performance (Table 2).

While identifying variation attributes is sufficient to cluster variant products, also extracting common attributes is critical to provide a complete description of the product group. To test VARM's ability to determine both common and variant attributes, catalog experts evaluated model predictions for both labels on a dataset with 10 variation groups. The evaluation showed that both version of VARM, $GenAI_{zero\_shot}$ and $GenAI_{RAG}$, can accurately predict common and variation attributes. Interestingly, providing information about the variation attributes using RAG not only improve performance for the variation attributes, but also the common attributes, suggesting that the generative LLM has a global understanding of the task and can generalize across the different rules (Table 3).

## 5 DISCUSSION

The recent developments in LLM technology has popularized its use with successful application to a multitude of use cases including e-commerce tasks [20]. Particularly, encoding LLMs are generally preferred for classification tasks or learning embeddings, while generative AI models are used for text generation tasks like summarizing or translation [3, 20, 28]. In this work, we capitalize on the respective advantages of encoding and generative LLMs to solve a new task for entity resolution aimed at identifying variant matches and variation attributes amongst e-commerce products.

Here, we show how we can learn variant product relationships leveraging the information present in website structures. Still, it will be worth exploring in future works how to adjust the granularity of these relationships so it can be applied across e-commerce sectors. For example, jewelry variation attributes could be material type at a coarse level of description, but for jewelry retailers the relevant variation attributes could be finer as ring size or gem type. Here, we showed that the strategy used to augment the dataset and generate negative examples was for model performance, suggesting that data augmentation methods could prove useful to define new types of product relationships [9, 26].

This work expands the traditional definition of ER to identify *variant relationships* and implemented a model to successfully identify this relationships amongst products. While it was only tested in e-commerce catalog applications, the new formulation and model strategy can be directly extended to other areas using ER, such as data curation or customer identification [2].

| Model | Recall (color, size) | Recall (all) |
|---|---|---|
| Heuristic | 70.65 | 70.28 |
| $GenAI_{zero\_shot}$ | 79.56 | 79.67 |
| $GenAI_{RAG}$ | **90.95** | **80.40** |

**Table 2: Variation attribute identification performance on the webpage-linked dataset sample.** Recall performance for models predicting color and size variation attributes, and all variation attributes.

| Model | Attribute type | Attribute # | Accuracy |
|---|---|---|---|
| $GenAI_{zero\_shot}$ | Common | 46 | 80.43 |
|  | Variation | 35 | 73.53 |
|  | All | 81 | 75.29 |
| $GenAI_{RAG}$ | Common | 46 | **84.78** |
|  | Variation | 35 | **74.29** |
|  | All | 81 | **79.01** |

**Table 3: Common and variation attribute identification performance.**

# A APPENDIX

| Model | Task | Parameter | Value |
|---|---|---|---|
| Claude instantV1 (GenAI$_{zero\_shot}$) | Product matching | max_tokens_to_sample | 30 |
| | | temperature | 0 |
| | | top_k | 100 |
| Claude3 Haiku (GenAI$_{zero\_shot}$, GenAI$_{RAG}$) | Attribute identification | max_tokens | 500 |
| | | temperature | 0 |
| | | top_p | 0.9 |

**Table 4: Generative model parameter settings.**

| Product | Product type | Brand | Variant attributes | URL |
|---|---|---|---|---|
| Razer Huntsman Mini - Linear Optical Switch - US - Mercury | Keyborad | Razer | Keyboard switch, Keyboard layout, Color/Design | https://www.razer.com/gaming-keyboards/Razer-Huntsman-Mini/RZ03-03390400-R3M1 |
| HyperX Alloy Origins - Mechanical Gaming Keyboard | Keyboard | HyperX | Switch, Model | https://hyperx.com/collections/gaming-keyboards/products/hyperx-alloy-origins-mechanical-gaming-keyboard?variant=42330451148957 |
| Linen-Blend Wide Strap Mini Dress | Dress | Abercrombie&Fitch | Color, Size, Length | https://www.abercrombie.com/shop/us/p/linen-blend-wide-strap-mini-dress-56256821 |
| Satin Slip Dress | Dress | Zara | Size | https://www.zara.com/us/en/satin-slip-dress-p08354475.html |
| Love Ring | Ring | Cartier | Size, Metal | https://www.cartier.com/en-us/jewelry/rings/love/love-ring-CRB4084800.html |
| Eternity solitaire ring | Ring | Swarovski | Color, Size | https://www.swarovski.com/en-US/p-M5697472/Eternity-solitaire-ring-Lab-grown-diamonds-1-2-ct-tw-Round-cut-14K-yellow-gold |

**Table 5: Example product listings on e-commerce platforms.**

## A.1 Variation product attributes

```
****** Variation product attributes ******
<product i>
brand = ...,
title = ...,
part number = ...,
model number = ...,
size = ...,
color = ...,
item package quantity = ...,
generic keyword = ...,
product category = ...,
product description = ...,
product type = ...,
...
<product i>
```

## A.3 Variant product classification prompt

```
****** Variant product classification prompt ******
You are an expert on products. Compare the details in the
two following products and determine if they refer to the
same or different products. The products are considered
the same product if all details match except for details like
color or size.
Return the answer strictly "yes" when they are the same
product or "no" when they are different. Also say, how
similar the two products are in a scale from 0 to 1? Where
0 means that the products are completely different and 1
they are the exact same.
Description of the first product: {product 1}
Description of the second product: {product 2}
First, respond "yes" or "no" and a similarity score between
0 and 1.
```

## A.2 Example variation product group attributes prediction outputs

```
{"Different": ["item_name", "item_id",
"item_package_weight", "color", "included_components",
"model_name", "size", "grip_size", "head_size"], "Same":
["brand", "product_type", "item_type"], "Reason": ["The
different attributes across the products are mainly related
to the specific details of each racket model, such as the
name, ID, images, weight, color, included components,
model name, size, grip size, and head size. These attributes
reflect the unique characteristics and specifications of
the different YONEX racket models. The same attributes
across the products indicate the common properties
shared by YONEX sport racket products, such as the
brand, product type and item type."]}
```

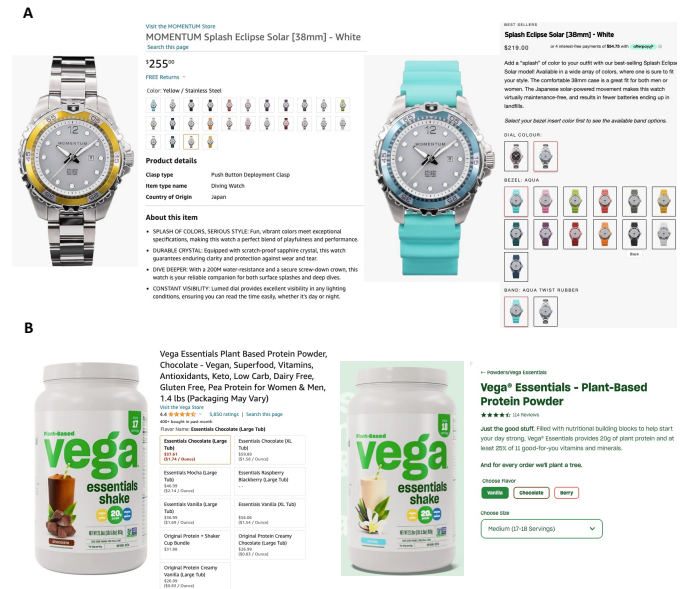

Figure 4: Example variant products identified by VARM. **A** Pair of watches belonging to the same variation group, regardless of difference in color, material or dial style, as shown by the shared listing on the webpage. **B** Pair of variant edible products where the variation attribute is *flavor*.

```
{"Different": ["color", "size", "item_name", "item_id",
"part_number", "generic_keyword"], "Same":
["age_range_description", "brand_value", "closure_type",
"material_composition", "item_type_keyword", "prod-
uct_type", "care_instructions"], "Reason": ["The color,
size, product names, item IDs, part numbers, and generic
keyword values are different across the two products to
represent different variations of the same product type.
However, the brand, closure type, material composition,
item type, product type, and care instructions are the same
across the two products as they are common attributes of
the product type."] }
```